# Solar-system tests of the inflation model with a Weyl term

**Wei-Tou Ni**[a,b]

[a]Shanghai United Center for Astrophysics (SUCA), Shanghai Normal University, 100 Guilin Road, Shanghai, 200234, China

[b]Center for Gravitation and Cosmology (CGC), Department of Physics, National Tsing Hua University, No. 101, Kuang Fu II Rd., Hsinchu, Taiwan, 30013, ROC

**Abstract**. Recently, there has been an interest in inflation and modified gravity with a Weyl term added to the general-relativistic action (N. Derulle, M. Sasaki, Y. Sendouda and A. Youssef, *JCAP*, **3**, 040 (2011)). In this paper we study empirical constraint on this modified gravity from solar-system experiments/observations. We first derive linearized equation of motion in the weak field limit and solve it for isolated system in the slow motion limit. We then use it to derive the light propagation equations, and obtain the relativistic Shapiro time delay and the light deflection in one-body central problem. Apply these results to the solar-system measurements, we obtain constraints on the Weyl term parameter $\gamma_W$; the most stringent constraint, which comes from the Cassini relativistic time delay experiment, is for $\gamma_W$ to be less than $1.5 \times 10^{-3}$ AU$^2$, or $|\gamma_W|^{1/2}$ less than 0.039 AU (19 s). Analysis of precision laboratory gravity experiments put further limit on the Weyl term parameter $\gamma_W$ to below the laboratory scale.

**Keywords**: solar-system tests of relativistic gravity, inflation, modified gravity, Weyl term, relativistic time delay

Contents

## 1 Introduction

Weyl [1] proposed the following action for the gravitational field from the point of view of scale invariance:

$$S_{\text{Weyl}} = -(\gamma_W/4\kappa)\int d^4x \, (-g)^{1/2} \, C_{\mu\nu\rho\sigma} C^{\mu\nu\rho\sigma}, \qquad (1.1)$$

Where $\gamma_W$ is a constant, $g$ is the determinant of the metric $g_{\mu\nu}$, and $C_{\mu\nu\rho\sigma}$ is the Weyl tensor,

$$C_{\mu\nu\rho\sigma} = R_{\mu\nu\rho\sigma} - (1/2)(g_{\mu\rho}G_{\nu\sigma} - g_{\mu\sigma}G_{\nu\rho} - g_{\nu\rho}G_{\mu\sigma} + g_{\nu\sigma}G_{\mu\rho}) - (R/3)(g_{\mu\rho}g_{\nu\sigma} - g_{\mu\sigma}g_{\nu\rho}). \qquad (1.2)$$

Since then, the Weyl action has been studied (i) due to its presence as quantum correction of various theories of quantum gravity [2]; (ii) as a phenomenological modification of general relativity [3]; (iii) in connection with particle theories [4].

To study inflation with a Weyl term, Deruelle, Sasaki, Sendouda and Youssef (DSSY) [5] considered the action

$$S = S_{\text{Hilbert-Einstein}} + S_{\text{scalar}} + S_{\text{Weyl}} = (1/2\kappa)\int d^4x \, (-g)^{1/2} R - (1/2)\int d^4x \, (-g)^{1/2} \, [\partial_\mu\varphi \, \partial^\mu\varphi + 2V(\varphi)]$$
$$- (\gamma_W/4\kappa)\int d^4x \, (-g)^{1/2} \, C_{\mu\nu\rho\sigma} C^{\mu\nu\rho\sigma}, \qquad (1.3)$$

The first term is the Hilbert-Einstein action; the second term is the scalar action; the third term is the Weyl action. In this paper, we use the units, $\kappa = 8\pi G_N$, $c = 1$ with $G_N$ the Newton gravitation constant unless otherwise specified, and adopt the (−+++) convention for the Minkowski metric $\eta_{\alpha\beta}$.[1] $\gamma_W$ is the coupling constant of the Weyl term (the last term in the action) and has dimension (length)$^2$. In [5], the Weyl term parameter $\gamma_W$ is denoted by $\gamma$. Since in the solar-system tests, $\gamma$ denotes one of the PPN parameters, here we use $\gamma_W$ to denote Weyl term parameter.

The second term $S_{\text{scalar}}$ represents scalar-field matter minimally coupled to the metric. To obtain solar-system gravity, a standard way [6] is to extend the scalar-field action $S_{\text{scalar}}$ to $S_{\text{matter}}$ to include all matter using minimal coupling principle (Einstein Equivalence Principle) [7], i.e.,

$$S = S_{\text{Hilbert-Einstein}} + S_{\text{matter}} + S_{\text{Weyl}}, \qquad (1.4)$$

where $S_{\text{matter}}(g_{\alpha\beta}, \Psi)$ is the special relativity action $S_{\text{special relativity}}(\eta_{\alpha\beta}, \Psi)$ with $\eta_{\alpha\beta}$ replaced by $g_{\alpha\beta}$ and ordinary derivation replaced by covariant derivation with respect to the Christoffel connection $\Gamma^\mu{}_{\nu\sigma}$ of the metric $g^{\mu\nu}$. $\Psi$ denotes all the matter fields including the scalar field in (1.3). Since the theory (1.4)

---

[1] Further conventions: $R^\mu{}_{\nu\rho\sigma} = \partial_\rho\Gamma^\mu{}_{\nu\sigma} - \partial_\sigma\Gamma^\mu{}_{\nu\rho} + \cdots$; $R_{\nu\sigma} = R^\mu{}_{\nu\mu\sigma}$; $R = g^{\mu\nu}R_{\mu\nu}$; $G_{\mu\nu} = R_{\mu\nu} - (1/2)g_{\mu\nu}R$. $\Gamma^\mu{}_{\nu\sigma}$ is the Christoffel connection of the metric $g^{\mu\nu}$. $R$ is the scalar curvature. Greek indices run from 0 to 3; Latin indices run from 1 to 3.

observes Einstein Equivalence Principle, it is in the class of metric theories.

Our paper is structured as follows. In section 2, we work out the linear approximation in the weak field. In section 3, we obtain the light propagation, the relativistic time delay and the light deflection in the linear approximation and in the weak-field slow-motion limit. In section 4, we examine constraints on the Weyl term parameter $\gamma_W$ from solar-system experiments. In section 5, we discuss our results and look into the accuracies of future solar-system measurements, their projected constraints, and the constraints on the Weyl term parameter $\gamma_W$ from laboratory experiments.

**2 Linear approximation in the weak field limit**

As in [5], from the variation of action (1.4), we obtain the equations of motion

$$G_{\mu\nu} - \gamma_W B_{\mu\nu} = \kappa\, T_{\mu\nu}, \tag{2.1}$$

where $B_{\mu\nu}$ is the Bach tensor [8]

$$B_{\mu\nu} = 2D^\rho D^\sigma C_{\mu\rho\nu\sigma} + G^{\rho\sigma} C_{\mu\rho\nu\sigma}, \tag{2.2}$$

and $T_{\mu\nu}$ is the stress-energy tensor derived from $S_{\text{matter}}(g_{\alpha\beta}, \Psi)$ in the usual way. $D^\rho$ denotes covariant derivation using the metric connection.

Any contraction of $C_{\mu\rho\nu\sigma}$ is zero. Using this property of $C_{\mu\rho\nu\sigma}$, the contraction of Bach tensor gives zero, i.e.,

$$B \equiv B_\mu{}^\mu = 0. \tag{2.3}$$

Contracting the equations of motion (2.1) and using (2.3), we have

$$R = -8\pi\, G_N T, \tag{2.4}$$

where $T \equiv T_\mu{}^\mu$. Substituting (2.4) into (2.1), we obtain the following equivalent equations of motion

$$R_{\mu\nu} - \gamma_W B_{\mu\nu} = 8\pi G_N[T_{\mu\nu} - (1/2)(g_{\mu\nu}T)]. \tag{2.5}$$

For weak field in the quasi-Minkowskian coordinates, we express the metric $g_{\alpha\beta}$ as

$$g_{\alpha\beta} = \eta_{\alpha\beta} + h_{\alpha\beta}, \qquad h_{\alpha\beta} \ll 1. \tag{2.6}$$

Since $h_{\alpha\beta}$ is a small quantity, we expand everything in $h_{\alpha\beta}$ and linearize the results to obtain the linear approximation. For linearization, we use the Minkowski metric $\eta_{\alpha\beta}$ to raise and lower indices without affecting the linearized results. The Riemann curvature tensor can be expressed as

$$R_{\alpha\beta\gamma\delta} = (1/2)(g_{\alpha\delta,\beta\gamma} + g_{\beta\gamma,\alpha\delta} - g_{\alpha\gamma,\beta\delta} - g_{\beta\delta,\alpha\gamma}) + g_{\mu\nu}(\Gamma^{\mu}{}_{\beta\gamma}\Gamma^{\nu}{}_{\beta\delta} - \Gamma^{\mu}{}_{\beta\delta}\Gamma^{\nu}{}_{\alpha\gamma}). \tag{2.7}$$

With linearization, we have

$$R_{\alpha\beta\gamma\delta} = (1/2)(h_{\alpha\delta,\beta\gamma} + h_{\beta\gamma,\alpha\delta} - h_{\alpha\gamma,\beta\delta} - h_{\beta\delta,\alpha\gamma}) + O(h^2), \tag{2.8}$$

$$R_{\alpha\gamma} = (1/2)(h_{\alpha\beta,\gamma}{}^{\beta} + h_{\beta\gamma,\alpha}{}^{\beta} - h_{\alpha\gamma,\beta}{}^{\beta} - h_{\beta}{}^{\beta}{}_{,\alpha\gamma}) + O(h^2), \tag{2.9}$$

$$R = h_{\alpha\beta,}{}^{\alpha\beta} - h_{\beta}{}^{\beta}{}_{,\alpha}{}^{\alpha} + O(h^2), \tag{2.10}$$

and

$$C_{\mu\nu\rho\sigma} = (1/2)(h_{\mu\sigma,\nu\rho} + h_{\nu\rho,\mu\sigma} - h_{\mu\rho,\nu\sigma} - h_{\nu\sigma,\mu\rho}) + (-1/2)(\eta_{\mu\rho}R_{\nu\sigma} - \eta_{\mu\sigma}R_{\nu\rho} - \eta_{\nu\rho}R_{\mu\sigma} + \eta_{\nu\sigma}R_{\mu\rho})$$
$$+ [(1/6)(\eta_{\mu\rho}\eta_{\nu\sigma}) - (1/6)(\eta_{\mu\sigma}\eta_{\nu\rho})](h_{\alpha\beta,}{}^{\alpha\beta} - h_{\beta}{}^{\beta}{}_{,\alpha}{}^{\alpha}) + O(h^2), \tag{2.11}$$

where $O(h^2)$ denotes terms of order of $h_{\alpha\beta}h_{\mu\nu}$ or smaller. Now we choose the harmonic gauge condition for $h_{\alpha\beta}$,

$$[h_{\alpha\beta} - (1/2)\eta_{\alpha\beta}h]_{,}{}^{\beta} = 0 + O(h^2, \gamma_W h), \text{ i.e., } h_{\alpha\beta,}{}^{\beta} = (1/2)h_{,\alpha} + O(h^2, \gamma_W h), \tag{2.12}$$

where $h$ outside of the $O(...)$ symbol is defined as $h \equiv h_{,\alpha}{}^{\alpha}$. In the $\gamma_W \to 0$ limit, this condition goes to the corresponding gauge condition in general relativity. Using the gauge condition (2.12), we calculate $D^{\nu}D^{\sigma}C_{\mu\nu\rho\sigma}$ and $B_{\mu\nu}$ to be

$$D^{\nu}D^{\sigma}C_{\mu\nu\rho\sigma} = (1/24)(\eta_{\mu\rho}h_{,\alpha}{}^{\alpha}{}_{\nu}{}^{\nu}) - (1/4)(h_{\mu\rho,\beta}{}^{\beta}{}_{\sigma}{}^{\sigma}) + (1/12)(h_{,\alpha}{}^{\alpha}{}_{\mu\rho}) + O(h^2, \gamma_W h), \tag{2.13}$$

$$B_{\mu\nu} = (1/12)(\eta_{\mu\nu}h_{,\alpha}{}^{\alpha}{}_{\rho}{}^{\rho}) - (1/2)(h_{\mu\nu,\beta}{}^{\beta}{}_{\sigma}{}^{\sigma}) + (1/6)(h_{,\alpha}{}^{\alpha}{}_{\mu\nu}) + O(h^2, \gamma_W h). \tag{2.14}$$

Using (2.9), (2.12) and (2.14), the equations of motion (2.5) in the linear approximation becomes

$$\{h_{\mu\nu} + \gamma_W[(1/6)\eta_{\mu\nu}h_{,\alpha}{}^{\alpha} - h_{\mu\nu,\sigma}{}^{\sigma} + (1/3)(h_{,\mu\nu})]\}_{,}{}^{\beta}{}_{\beta} = -16\pi G_N[T_{\mu\nu} - (1/2)(\eta_{\mu\nu}T)] + O(h^2, \gamma_W{}^2 h). \tag{2.15}$$

Choosing retarded solution and integrating as in classical electrodynamics/general relativity, we have (restoring factors of $c$)

$$h_{\mu\nu} + \gamma_W[(1/6)\eta_{\mu\nu}h_{,\alpha}{}^\alpha - h_{\mu\nu,\sigma}{}^\sigma + (1/3)(h_{,\mu\nu})] = [(4G_N)/(c^4)]\int\{[T_{\mu\nu} - (1/2)(\eta_{\mu\nu}T)]/r\}_{\text{retarded}}(d^3\underline{x}') + O(h^2, \gamma_W{}^2 h), \quad (2.16)$$

where $r = |\underline{x}-\underline{x}'|$. To solve this equation, we expand $h_{\mu\nu}$ in terms of $\gamma_W$ as:

$$h_{\mu\nu} = h^{(0W)}{}_{\mu\nu} + h^{(1W)}{}_{\mu\nu} + O(\gamma_W{}^2 h), \quad (2.17)$$

where $h^{(0W)}{}_{\mu\nu}$ is zeroth order in $\gamma_W$, i.e., of order $O(h)$ and $h^{(1W)}{}_{\mu\nu}$ first order in $\gamma_W$ i.e., of order $O(\gamma_W h)$. Substituting (2.17) into (2.16) and solving iteratively, we obtain

$$h^{(0W)}{}_{\mu\nu} = [(4G_N)/(c^4)]\int\{[T_{\mu\nu} - (1/2)(\eta_{\mu\nu}T)]/r\}_{\text{retarded}}(d^3x') \ (= h_{\mu\nu}{}^{GR}, \text{ and } h^{(0W)}{}_{\mu\nu,\alpha}{}^\alpha = O(h^2)), \quad (2.18)$$

$$h^{(1W)}{}_{\mu\nu} = (-1/3)[\gamma_W h^{(0W)}{}_{,\mu\nu}]. \quad (2.19)$$

We notice that $h_{\mu\nu}{}^{(0W)}$ is the same as in general relativity and the gauge condition implies

$$[h_{\alpha\beta}{}^{(0W)} - (1/2)\eta_{\alpha\beta}h^{(0W)}]{,}^\beta = O(h^2). \quad (2.20)$$

Using (2.18-2.20) to calculate $[h^{(1W)}{}_{\alpha\beta} - (1/2)\eta_{\alpha\beta}h^{(1W)}]{,}^\beta$, we have

$$[h^{(1W)}{}_{\alpha\beta} - (1/2)\eta_{\alpha\beta}h^{(1W)}]{,}^\beta = -(1/6)\gamma_W h^{(0W)}{}_{,\alpha\beta}{}^\beta = \gamma_W O(h^2) = O(\gamma_W h^2). \quad (2.21)$$

With (2.20) and (2.21), our solution satisfies the gauge condition (2.12) to order $O(h^2, \gamma_W h^2)$, i.e.,

$$[h_{\alpha\beta} - (1/2)\eta_{\alpha\beta}h]{,}^\beta = O(h^2, \gamma_W h^2), \quad (2.22)$$

and the solution $h_{\mu\nu}$ is

$$h_{\mu\nu} = [(4G_N)/(c^4)]\int\{[T_{\mu\nu} - (1/2)(g_{\mu\nu}T)]/r\}_{\text{retarded}}(d^3x') - (1/3)\gamma_W h^{(0W)}{}_{,\mu\nu} + O(h^2, \gamma_W h^2). \quad (2.23)$$

From now on, we impose slow motion condition, in addition to weak field, i.e.,

$$U/c^2 = O(v^2/c^2); \ U_{,ij}/c^2 = (1/L^2)\,O(v^2/c^2); \ U_{,0i}/c^2 = (1/L^2)\,O(v^3/c^3); \ U_{,00}/c^2 = (1/L^2)\,O(v^4/c^4), \quad (2.24)$$

where $L$ is a typical length scale of the system [7]. The solution of (2.18) and (2.19) then becomes

$$h^{(0W)}{}_{\mu\nu} = h^{GR}{}_{\mu\nu} = 2(U/c^2)\delta_{\mu\nu} + O(v^3/c^3); \ h^{(0W)} \equiv h^{(0W)}{}_\mu{}^\mu = 4\,(U/c^2) + O(v^3/c^3), \quad (2.25)$$

$$h^{(1W)}{}_{\mu\nu} = -(4/3)(\gamma_W U_{,\mu\nu}/c^2) + O(\gamma_W v^3/c^3), \quad (2.26)$$

where $h^{GR}_{\mu\nu}$ is the general relativity value and $U$ is the Newtonian potential. For point mass or outside spherical Sun,

$$U = (G_N M/c^2)(1/r); \quad r = (x^2 + y^2 + z^2)^{1/2}. \tag{2.27}$$

Hence, in the DSSY theory [5] with a Weyl term,

$$h_{\mu\nu} = (2U/c^2)\delta_{\mu\nu} - (4/3)(\gamma_W U_{,\mu\nu}/c^2) + O(\gamma_W v^3/c^3). \tag{2.28}$$

The Weyl term gives fast variation with distance to the source. This is a general property of equations of motion with higher derivatives.

**3 Light propagation, relativistic time delay and deflection**

In this section, we derive the light propagation equation in the weak field limit for the modified gravity theory with the Weyl term (1.4). Let $\underline{r} = \underline{r}(t)$ be the light trajectory where $\underline{r}(t) = (x(t), y(t), z(t))$ is a 3-vector. Light propagation follows null geodesics of the metric $g_{\alpha\beta}$, and its trajectory $\underline{r}(t)$ satisfies

$$0 = ds^2 = g_{\alpha\beta}dx^\alpha dx^\beta = (-1 + h_{00})c^2 dt^2 + 2h_{0i}cdx^i dt + (\eta_{ij} + h_{ij})dx^i dx^j, \tag{3.1}$$

where we have used eq. (2.6) in the weak field limit with $h_{\alpha\beta} \ll 1$.

In the Minkowski approximation, the light trajectory can be approximated by

$$dx^i/dt = (dx^i/dt)^{(0)i} + O(h) = cn^{(0)i} + O(h), \text{ with } \sum_i (n^{(0)i})^2 = 1, \tag{3.2}$$

where $n^{(0)i}$ are constants. In the Post-Minkowski approximation, we express $dx^i/dt$ as

$$dx^i/dt = cn^{(0)i} + cn^{(1)i} + O(h^2), \tag{3.3}$$

where $n^{(1)i}$ is a function of trajectory and of the order of O($h$). Substituting (3.3) into (3.1) and dividing by $dt^2$, we have

$$0 = (-1 + h_{00})c^2 + 2h_{0i}c(dx^i/dt) + |d\underline{r}/dt|^2 + h_{ij}[(dx^i/dt)(dx^j/dt)] \tag{3.4a}$$
$$= (-1 + h_{00})c^2 + 2h_{0i}c(cn^{(0)i} + cn^{(1)i}) + c^2\sum_{i=1}^{3}(n^{(0)i} + n^{(1)i})^2 + h_{ij}n^{(0)i}n^{(0)j}c^2 + O(h^2) = 0. \tag{3.4b}$$

Simplifying (3.4b), we have

$$\sum_{i=1}^{3} n^{(0)i} n^{(1)i} = -(1/2)(h_{00} + 2h_{0i}n^{(0)i} + h_{ij}n^{(0)i}n^{(0)j}) + O(h^2). \quad (3.5)$$

Solving for $|d\underline{r}/dt|$ in (3.4a), we obtain the light propagation equation to $O(h)$:

$$\begin{aligned}|d\underline{r}/dt| &= c[(1 - h_{00} - 2h_{0i}n^{(0)i} - h_{ij}n^{(0)i}n^{(0)j} + O(h^2)]^{1/2} \\ &= c[1 - (1/2)h_{00} - h_{0i}n^{(0)i} - (1/2)h_{ij}n^{(0)i}n^{(0)j} + O(h^2)].\end{aligned} \quad (3.6)$$

From (3.6), we calculate the relativistic Shapiro time delay $t_S$ [9] as

$$t_S = (1/c)\int |d\underline{r}|\,[1 + (1/2)h_{00} + h_{0i}n^{(0)i} + (1/2)h_{ij}n^{(0)i}n^{(0)j} + O(h^2)]. \quad (3.7)$$

In the theory with a Weyl term, choosing the z-axis along the initial light propagation direction, i.e., $n^{(0)i} = (0, 0, 1)$, and using (2.28) for a slow-motion system, we have

$$\begin{aligned}\Delta t_S = \int dt &= (1/c)\int dz[1 + 2U - (2/3)(\gamma_W U_{,zz}) + O(h^2)] = \Delta t^N + \Delta t_S^{GR} - (2/3)(\gamma_W U_{,z})|_{z1}^{z2} \\ &= (1/c)(z_2 - z_1) + 2(GM/c^3)\,ln\{[(z_2^2 + b^2)^{1/2} + z_2]/[(z_1^2 + b^2)^{1/2} + z_1]\} \\ &\quad + (2/3)[\gamma_W(GM/c^2)][(z_2/r_2^3) - (z_1/r_1^3)] + O(h^2), \quad (z_1 < 0, z_2 > 0),\end{aligned} \quad (3.8)$$

where the first term is the Newtonian travel time, the second term is the general-relativistic Shapiro time delay, the third term is the additional Shapiro time delay due to the Weyl term, and $b$ is the impact parameter of light propagation.

The geodesic equation for light and for test particle in general relativity and in the metric theories of gravity

$$d^2x^\mu/d\lambda^2 + \Gamma^\mu_{\sigma\rho}(dx^\sigma/d\lambda)(dx^\rho/d\lambda) = 0, \quad \lambda: \text{affine parameter} \quad (3.9)$$

can be cast in the form

$$d(g_{\mu\nu}dx^\nu/d\lambda)/d\lambda = (1/2)\,g_{\sigma\rho,\mu}(dx^\sigma/d\lambda)(dx^\rho/d\lambda). \quad (3.10)$$

Integrating, we obtain

$$g_{\mu\nu}dx^\nu/d\lambda = (1/2) \int [g_{\sigma\rho,\mu}(dx^\sigma/d\lambda)(dx^\rho/d\lambda)]\,d\lambda. \quad (3.11)$$

To obtain light deflection angle in a weak gravitational field of the Sun or other source, we choose x-axis in the initial light (photon) propagation direction, y-axis in the plane spanned by the Sun or other gravitational source and the light trajectory, and the sense of the y-axis is in the direction of the trajectory. From (3.11), we obtain

$$g_{0y} + dy/dt = (1/2) \int (h_{00,y} + h_{xx,y} + 2h_{0x,y}) \, dt + O(h^2). \tag{3.12}$$

Solving for $dy/dt$ in (3.12), substituting in (2.28) and simplifying, we obtain

$$dy/dt = -g_{0y} + (1/2) \int (h_{00,y} + h_{xx,y} + 2h_{0x,y}) \, dt + O(h^2) = \int \{(2/c^2)U_{,y} - (2/3)\gamma_W U_{,xxy}/c^2\} \, dt$$
$$= \{-(2/c^2) G_N M b/(x^2+b^2)^{1/2} - (2/3)\gamma_W U_{,xy}/c^2\}|_{x=x_1} - \{(2/c^2) G_N M x/(x^2+b^2)^{1/2} - (4/3)\gamma_W U_{,xy}/c^2\}|_{x=x_0}, \tag{3.13}$$

where

$$U_{,xy} = 3 G_N M \, x \, y / (x^2+b^2)^{5/2}. \tag{3.14}$$

If the source of light (or radio signal) is far away, the second term in the second curly brackets of (3.13) can be dropped and we have

$$\Delta \varphi_{\text{deflection}} = -2 \, (G_N M/c^2 b)(\sin\theta_1 - \sin\theta_0) - 2 \, (G_N M /c^2 b)(b^2/r_1^2)(x_1/r_1)(\gamma_W/r_1^2), \tag{3.15}$$

for the deflection angle $\Delta\varphi_{\text{deflection}}$, where $\theta_0$ ($\theta_1$) is the angle between the position vector of the light emitter (observer) and $x$-axis. For close impact, $b \ll r_1$, we have

$$\Delta\varphi_{\text{deflection}} = -2 \, (G_N M/c^2 b)(\sin\theta_1 + 1) - 2 \, (G_N M /c^2 b)(b^2/r_1^2)(\gamma_W/r_1^2). \tag{3.16}$$

with $\sin\theta_1 \approx 1$.

**4 Constraint on the Weyl term parameter $\gamma_W$ from solar-system experiments**

In this section, we use the solar-system measurements [10-12] of the relativistic time delay and light deflection to constrain the Weyl term parameter $\gamma_W$. The error of these experiments is usually quoted in terms of PPN parameter $\gamma$. Since the effect is proportional to $(1 + \gamma)$, the agreement with general relativistic effect is within half of the value of the error for $\gamma$.

The most precise experiment for relativistic time delay measurement in the solar system is the Cassini time delay experiment [10, 13]. The Cassini experiment was carried out between 6 June and 7 July 2002, when the spacecraft was on its way to Saturn, around the time of a solar conjunction. The conjunction—at which the spacecraft (at a geocentric distance of 8.43 AU), the Sun and the Earth were almost aligned, in this order—occurred on 21 June 2002, with a minimum impact parameter $b$ of 1.6 $R_S$ (solar radius), and no occultation. At this time, there is a maximum two-way general relativistic Shapiro time delay of 260 μs according to second term in (3.8). The variation of this time delay, through 18 Doppler frequency measurements along spacecraft passages was verified to $0.5 \times (2.1 \pm 2.3)$

× $10^{-5}$. Roughly speaking, the room for the third term with $\gamma_W$ is less than $2.2 \times 10^{-5}$ of the Shapiro time delay. Therefore, $|\gamma_W|$, the absolute value of $\gamma_W$, is constrained to

$$|\gamma_W| \leq 1.5 \times 10^{-3} \text{ AU}^2, \text{ or } |\gamma_W|^{1/2} \leq 0.039 \text{ AU } (= 19 \text{ s}). \tag{4.1}$$

To be more precise, a fitting of original data to equation (3.8) is needed. Since the Weyl term effect gives faster variation in the Doppler frequency shift than the general relativistic Shapiro effect, actual fitting would give better constraint on $|\gamma_W|$.

Let us now work out the constraint set from the light (radio wave) deflection experiments. In 2004, Shapiro, Davis, Lebach and Gregory [11] used very-long-baseline interferometry (VLBI) to measure the deflection by the Sun of radio waves emanating from distant compact radio sources. They determined his bending using a large geodetic VLBI data and obtained the value $0.5 \times [1 + (0.99983 \pm 0.00045 \text{ (estimated standard error)})]$ in terms of the general relativistic value.

Fomalont, Kopeikin, Lanyi and Benson [14] used the Very Long Baseline Array (VLBA) at 43, 23, and 15 GHz to measure the solar gravitational deflection of radio waves among four radio sources during an 18 day period in October, 2005. Using phase-referenced radio interferometry to fit the measured phase delay to the propagation equation, the deflection is determined to be $0.5 \times [1 + (0.9998 \pm 0.0003 (68\% \text{ confidence level}))]$ of the general relativity value.

Lambert and Le Poncin-Lafitte [15] analyzed the geodetic VLBI observations recorded since 1979 using several strategies and various data sets covering different time spans, and arrived at the conclusion that $\gamma$ is unity within $2 \times 10^{-4}$, i.e., the agreement to general relativistic effect is $1 \times 10^{-4}$.

From these 3 experiments, the room for the deviation from general relativistic light deflection is less than $1 \times 10^{-4}$. Therefore, from equation (3.16), we have

$$2 \times |(b^2/1\text{AU}^2)(\gamma_W/1\text{AU}^2)| \leq 10^{-4}. \tag{4.2}$$

In the observations [11, 14, 15], the smaller impact parameters are around $10\, R_S$. Hence, we can readily obtain the following constraint:

$$|(\gamma_W/1\text{AU}^2)| \leq 0.02, \text{ or } |\gamma_W|^{1/2} \leq 0.4 \text{ AU } (= 200 \text{ s}), \tag{4.3}$$

from these radio wave deflection experiments. Since the Weyl term effect on the light deflection is proportional to the impact parameter $b$, a detailed analysis is likely to give better constraint.

**5 Discussion and Outlook**

We have studied empirical constraint on the DSSY modified gravity and inflation model from solar-system experiments/observations. After deriving the linearized equation of motion in the weak field limit and solving it in the slow motion approximation, we derived the light propagation equations, the relativistic time delay and the light deflection. Applying these results to the solar-system measurements, we obtained constraints on the absolute value of the Weyl term parameter $\gamma_W$ to be less than $1.5 \times 10^{-3}$ AU$^2$, or $|\gamma_W|^{1/2}$ less than 0.039 AU, from the Cassini relativistic time delay experiment. This is in contrast to Pireaux's analysis of Weyl's gravity [1] putting a constraint on the length scale of the linear parameter to larger than about $10^{19}$ m (i.e., $0.7 \times 10^8$ AU) [3].

There are a few solar-system missions that can improve on the test of relativistic gravity. GAIA (Global Astrometric Interferometer for Astrophysics) is an astrometric mission concept aiming at the broadest possible astrophysical exploitation of optical interferometry using a modest baseline length (3 m) [16]. GAIA is planned to be launched in 2013. At the present study, GAIA aims at limit magnitude 21, with survey completeness to visual magnitude 19-20, and proposes to measure the angular positions of 35 million objects (to visual magnitude V = 15) to 10 μas accuracy and those of 1.3 billion objects (to V = 20) to 0.2 mas accuracy. The observing accuracy of V = 10 objects is aimed at 4 μas. To increase the weight of measuring the relativistic light deflection parameter $\gamma$, GAIA is planned to do measurements at elongations greater than 35° (as compared to essentially 47° for Hipparcos) from the Sun. With all these, a simulation shows that GAIA could measure $\gamma$ to $1 \times 10^{-5}$ - $2 \times 10^{-7}$ accuracy [17].

NASA's MESSENGER spacecraft has been in orbit around the planet Mercury since March 2011 [18]. As a joint mission of Europe and Japan to Mercury, BepiColombo will set off in 2015 and arrive at Mercury in January 2022 for its 1 year nominal mission, with a possible 1-year extension [19]. A simulation predicts that the determination of the PPN parameter $\gamma$ can reach $2 \times 10^{-6}$ through relativistic Shapiro measurement of radio waves [20]. ASTROD I mission is a dedicated relativity-test mission concept using optical ranging to a single spacecraft from Earth and proposing to measure the PPN parameter $\gamma$ to $3 \times 10^{-8}$ through relativistic Shapiro measurement of laser signals [21, 22].

These future missions would improve the measurement of the PPN parameter $\gamma$ by 1-3 orders of magnitude. Hence, the measurement/constraint of $\gamma_W$ would be improved by 1-3 orders of magnitude also.

From (2.28), the Weyl term modification of gravity in terms of strength of Newtonian gravity is of magnitude $(2/3)\gamma_W U_{,\mu\nu}/U$. In the laboratory gravitation experiments, the Weyl term will give significant deviations to Newtonian gravity if $|\gamma_W|^{1/2}$ is of the order of magnitude of experimental scale. Analysis of various precision laboratory gravitation experiments in the scale range of millimeter to meter would put further limit on the Weyl term parameter $|\gamma_W|^{1/2}$ of the order 1 mm or less.

The smallness of the Weyl term parameter in DSSY theory makes the term ineffective in the cosmological setting.

Very recently, Deruelle *et al.* [23] have shown that the ghost degrees of freedom of Einstein gravity with a Weyl term can be eliminated by a simple mechanism that invokes local Lorentz symmetry breaking via *chronon* scalar field. They have also shown how the mechanism works in a cosmological setting. It would be worthwhile to study the solar-system constraints and laboratory constraints on the modified Weyl term parameter of this theory.

In the linear approximation, the DSSY theory [5] does not admit monopole radiation. In a previous study, we found that both Yang's gravitational field equation [24] and the $C^{\mu\nu\rho\sigma} = 0$ equation admits monopole radiations [25]. The metric

$$ds^2 = [c_0 + f(r-t)/r + g(r+t)/r]^2(-dt^2 + dr^2 + r^2 d\Omega^2), \tag{5.1}$$

is an exact vacuum solution with arbitrary functions $f$ and $g$ of both equations. However this metric is not a solution of the DSSY theory [5] unless $f$ and $g$ vanish. It would be worthwhile to see whether the new version of DSSY theory without ghosts [23] has monopole solutions.

**Acknowledgements**

I thank Natalie Deruelle and Misao Sasaki for helpful discussions during the ICGCA 10 Conference and the 2012 APS School and Workshop. I also thank the National Science Council (Grants No. NSC100-2119-M-007-008 and No. NSC100-2738-M-007-004) for supporting this work in part.